\documentclass{aa}  

\usepackage{graphicx}   
\usepackage{amsmath}    
\usepackage{amssymb}    
\usepackage{multirow}
\usepackage{float}
\usepackage{txfonts}
\usepackage[colorlinks=true,citecolor=blue, linkcolor=blue, urlcolor=blue]{hyperref}
\begin{document}

   \title{Repeating X-ray bursts: Interaction between a neutron star and clumps partially disrupted from a planet}


   \author{Abdusattar Kurban\inst{1,2,3}
        \and
        Xia Zhou\inst{1,2,3}
        \and
        Na Wang\inst{1,2,3}
        \and
        Yong-Feng Huang\inst{4,5,1}
        \and
        Yu-Bin Wang\inst{6}
        \and
        Nurimangul Nurmamat\inst{4}
}

\institute{Xinjiang Astronomical Observatory, Chinese Academy of Sciences, Urumqi 830011, Xinjiang, People's Republic of China\\
        \email{akurban@xao.ac.cn (AK); na.wang@xao.ac.cn (NW); zhouxia@xao.ac.cn (XZ)}
        \and
        Key Laboratory of Radio Astronomy, Chinese Academy of Sciences, Urumqi 830011, Xinjiang, People's Republic of China
        \and
        Xinjiang Key Laboratory of Radio Astrophysics, Urumqi 830011, Xinjiang, People's Republic of China
        \and
        School of Astronomy and Space Science, Nanjing University, Nanjing 210023, People's Republic of China
        \and
        Key Laboratory of Modern Astronomy and Astrophysics (Nanjing University), Ministry of Education, Nanjing 210023, People's Republic of China
        \and
        School of Physics and Electronic Engineering, Sichuan University of Science \& Engineering, Zigong 643000, People's Republic of China
        }

   \date{Received Month xx, yyyy; accepted Month xx, yyyy}

 
  \abstract
   {Repeating X-ray bursts from the Galactic 
   magnetar SGR 1806-20 have been observed with a period of 398 days. 
   Similarly, periodic X-ray bursts from SGR 1935+2154 with a period of 238 days 
   have also been observed. Here we argue that 
   these X-ray bursts could be produced by the interaction of 
   a neutron star (NS) with its planet in a highly elliptical orbit.
   The periastron of the planet is very close to the NS,
   so it would be partially disrupted by the tidal
   force every time it passes through the periastron.
   Major fragments generated in the process will fall onto the
   NS under the influence of gravitational perturbation.
   The collision of the in-falling fragments
   with the NS produces repeating X-ray bursts.
   The main features of the observed X-ray bursts, 
   such as their energy, duration, periodicity, and
   activity window, can all be explained in our 
   framework.} 

   \keywords{Stars: neutron -- Planet-star interactions -- X-rays: bursts}

   \maketitle



\section{Introduction} \label{sec:intro}

Magnetars are highly magnetized neutron stars
\citep[NSs;][]{Thompson1995MNRAS} that are
characterized by recurrent emission of short-duration
bursts (a few milliseconds to seconds) in soft $\gamma$-rays and/or hard X-rays, with burst energies in the range
$10^{35}$ to $10^{46}$ erg (see the review articles \citealt{Turolla2015} and
\citealt{Kaspi2017ARAA} and reference therein).
More than 30 magnetars have been observed to date,\footnote{\url{http://www.physics.mcgill.ca/~pulsar/magnetar/main.html}}
and they show diverse burst properties \citep{Olausen2014ApJS}.

Many models have been proposed to account for magnetar burst activities.
According to their energy sources, these models can be divided into
two categories: internal and external mechanisms.
In the internal mechanism models, the burst energy mainly comes from the magnetic field,
whose decay leads to a rupture of the crust and rapid magnetic
reconnection, which releases a large amount of energy and
causes particle acceleration and radiation in the magnetosphere
\citep{Thompson1995MNRAS,Lyutikov2003MNRAS,Gill2010MNRAS,Turolla2015}.

On the other hand, in the external mechanism models,
the energy comes from the interaction between the compact stars -- for example NSs or strange stars (SSs) -- and external material.
The bursts could also be due to the impact of
massive comet-like objects with a NS \citep{Katz1996ApJ, Chatterjee2000ApJ, Marsden2001ApJ}
or a SS \citep{Zhang2000ApJ, Usov2001PhRvL, Ouyed2011MNRAS}. 
We note that in the models the origin and location of these small bodies can be
different. They can come from a fossil disk formed after the birth
of the NS and/or SS \citep{Chatterjee2000ApJ, Marsden2001ApJ} or from the comet clouds
around the central star (like the Oort cloud in the Solar System; \citealt{Zhang2000ApJ}).
They can even be formed during a quark nova explosion, an explosive phenomenon
that happens when a NS is converted into a SS due to spin-down
or accretion processes \citep{Ouyed2011MNRAS}. 
We note that both NSs and SSs
could be involved in the bursts and super-Eddington activities \citep[e.g.,][]{Katz1996ApJ, Geng2021Innov}.

At present, it is not clear which of the various mechanisms is in operation.\ The richness of the phenomenology indicates that a combination of mechanisms are responsible \citep{Turolla2015}.

Recent studies show that two soft gamma-ray repeaters (SGRs), SGR 1806-20
and SGR 1935+2154, may have periodic burst activities.
SGR 1806-20 was initially known as GB790107 and was identified
as a gamma-ray repeater by \citet{Laros1986Natur}.
It has a surface magnetic field of $ B_{\star} = 2\times 10^{15}$ G \citep{Olausen2014ApJS}
and a spin period of $ P_{\rm spin} = 7.55$ s \citep{Woods2007ApJ}.
\citet{Zhang2021ApJ} analyzed the arrival time of more than 3000 short bursts from SGR 1806-20 detected by different telescopes \citep{Ersin2000ApJ, Prieskorn2012ApJ,bayrak2017ApJS}
and found a possible period of about 398 days for the source.

SGR 1935+2154 was discovered by \citet{Stamatikos2014GCN}.
The surface magnetic field of this source is $ B_{\star} = 2.2\times 10^{14}$ G,
and the spin period is $ P_{\rm spin} = 3.245 $ s \citep{Israel2016MNRAS}.
After analyzing the arrival time of more than 300 X-ray bursts
from SGR 1935+2154, \citet{Zou2021ApJ} argued that these bursts
exhibit a period of $\sim 238$ days, with a 150 day active window (63.3\% of the period).
However, \citet{Xie2022MNRAS} revisited the periodicity
of SGR 1935+2154 with an updated sample and found the period to
be 126.88 days. Further investigation into the periodical activities
of SGR 1935+2154 is still urgently needed.

The periodic activities of SGRs may be caused by the
precession of a NS \citep{Zhang2021ApJ, Zou2021ApJ}, when the emission
region periodically sweeps our line of sight. In this framework,
a burst occurs when the crust fractures or a magnetic reconnection happens.
The underlying cause of these activities is likely the evolution of the magnetic field,
which itself depends dramatically on a large number of degrees
of freedom for the large-scale magnetospheric configuration \citep{Turolla2015}.
Details concerning the triggering mechanism, the efficient conversion of magnetic
energy into thermal energy, and the transportation of the thermal energy to the
surface are still largely uncertain \citep{Kaspi2017ARAA,Gourgouliatos2018}.

On the other hand, a wide variety of transients can be produced during the tidal
disruption processes. A tidal disruption happens when an object
gets too close to its host star, triggering so-called
tidal disruption events (TDEs). This could occur in any system if
the tidal disruption condition is satisfied \citep{Hills1975Natur,Rees1988Natur}.
    Several classes of TDEs with emissions in 
        the optical/UV, soft X-rays, hard X-rays, and gamma-rays have been
        reported (see \citealt{Gezari2021ARAA} and references therein).
        It was recently reported by \cite{Cendes2023arXiv230813595C}
        that many TDEs exhibit late-time radio emission, often months to years
        after the initial optical/UV flare. It has also been argued that
repeating transient events, for example AT 2018fyk \citep{Wevers2023ApJL},
        AT 2020vdq \citep{Somalwar2023arXiv231003782S}, ASASSN-14ko 
        \citep{Payne2021ApJ, Cufari2022ApJ, Huang2023ApJ}, RX J133157.6-324319.7 \citep{Malyali2023MNRAS},
        eRASSt J045650.3-203750 \citep{Liu2023AA}, and 
        Swift J023017.0+283603 \citep{Evans2023NatAs}, may be due to a repeating partial disruption.

    Various emission mechanisms have been proposed to
        explain these observations. It is widely believed
        that the interaction between an outflow and the dense
        circumnuclear medium produces the radio emission in TDEs
        \citep{Cendes2022ApJ, Cendes2023arXiv230813595C}.
        For the UV/optical emission, however, we still lack a clear picture.
        Some models include the radiation from shocks generated 
        in stream--stream collisions \citep{Piran2015ApJ,Shiokawa2015ApJ,Coughlin2022ApJ}
        or the reprocessing of X-rays in either outflowing
        \citep[e.g.,][]{Strubbe2009MNRAS,Metzger2016MNRAS,Roth2018ApJ} 
        or static materials \citep{Loeb1997ApJ,Guillochon2014ApJ,Roth2016ApJ}.
        The soft X-ray emission is thought to come from a compact accretion
        disk \citep[e.g.,][]{Komossa1999AA, Auchettl2017ApJ}.
        The hard X-rays in some sources may be related to
        Comptonization in a relativistic outflow and
        amplification due to the beaming effect
        \citep[e.g.,][]{Bloom2011Sci, Cenko2012ApJ, Pasham2023NatAs},
        while in others it is likely due to sub-Eddington
        accretion and the formation of a corona 
        \citep[e.g.,][]{Wevers2019MNRAS, Lucchini2022MNRAS}.

It is interesting to note that the tidal disruption of a planet by its
compact host star under
various circumstances has been investigated by many authors. It may be related, for example, to the pollution of
the white dwarf atmosphere by heavy elements \citep{Malamud2020a, Malamud2020b}.
It may be connected with the close-in strange quark planetary
systems \citep{Geng2015, Huang2017ApJ, Kuerban2020ApJ} or be a mechanism
that produces some kinds of fast radio bursts \citep{Kurban2022ApJ} or
gamma-ray bursts \citep{Colgate1981ApJ, Campana2011Natur}.
It may also lead to glitch and/or anti-glitches \citep{Huang2014, Yu2016RAA}
and X-ray bursts \citep{Huang2014, Geng2020, Dai2016, Dai2020b}.
Recently, the dynamics of the clumps generated during the partial
disruption of a planet by its compact host was studied in detail \citep{Kurban2023MNRAS}.
It is found that the clumps could lose their angular momentum quickly
and fall toward the central star on a short timescale.

        Repeating partial tidal disruption generally requires the 
        planet to be in a highly elliptic orbit with a small pericenter distance
        of $\sim 10^{11}$ cm. How such a planet can be formed needs to be clarified. 
        It is well known that NSs are born during the death of massive stars, at  
        which time close-in planets would be engulfed by the expanding envelope of the
        star and destroyed. To date, however, more than 20 pulsars 
    accompanied by the candidate planets have been observed (see the Extrasolar Planets 
        Encyclopaedia\footnote{\url{https://exoplanet.eu/catalog/}} and \citealt{Schneider2011AA}).
        Several scenarios have been proposed to explain the existence 
        of pulsar planets. Firstly, a planet may survive the
        red-giant stage of a massive star \citep{Bailes1991Natur} and
        acquire an eccentric orbit thanks to the kick of the newborn NS produced in an asymmetric supernova explosion \citep{Bhattacharya1991PhR,Thorsett1993ApJ}.
        Secondly, the planet could be formed in the fallback disk of the supernova  
        \citep{Currie2007ApJ,Hansen2009ApJ}, or due to the destruction of a
        binary companion \citep{Martin2016ApJ}. Thirdly, a NS could 
        capture a passing-by planet most likely moving in a 
        highly eccentric orbit \citep{Podsiadlowski1991Natur}. For example,
        the circumbinary planet around PSR B1620-26 in a globular cluster 
        is thought to form in this way \citep{Sigurdsson2003Sci}.
        Although the orbits of currently detected pulsar planets generally
        have a small ellipticity, high-eccentricity pulsar-planet systems
        can be formed from dynamical processes such as capturing
        \citep[e.g.,][]{Goulinski2018MNRAS, Kremer2019ApJ, Cufari2022ApJ}, scattering
        \citep[e.g.,][]{Hong2018ApJ, Carrera2019AA}, or the Kozai-Lidov mechanism
        \citep[e.g.,][]{Kozai1962, Lidov1962, Naoz2016ARAA}.
    They are more likely to be found in globular 
        clusters\footnote{\url{https://www3.mpifr-bonn.mpg.de/staff/pfreire/GCpsr.html}}.
        The current non-detection of highly eccentric planets around pulsars
        may be due to their large distances from us as well as various other
        observational biases.

Motivated by these studies, we propose a new model to explain
the periodic X-ray bursts of SGRs. In our scenario, a planet is orbiting
around a magnetar in a highly eccentric close-in orbit. Every time the planet
passes through the periastron, it is partially disrupted by the
tidal force from its host. The clumps generated in this way fall
onto the magnetar and produce X-ray bursts. The observed periodicity
of SGRs can be explained by our model.

The structure of this paper is as follows.
In Section \ref{sec:model-para} the basic picture
of our model is described, and key parameters of the planet and clumps
disrupted from the planet are introduced.
In Section \ref{sec:xrb} the main properties of the X-ray bursts
are predicted, such as the energetics, durations, periodicity, and
activity window. In Section \ref{sec:comparison} we compare our
theoretical results with observations.
Finally, Section \ref{sec:conclusion} presents our
conclusion and some brief discussions.

\section{Model and parameters}\label{sec:model-para}

\subsection{Model}\label{subsec:model}

\begin{figure*}
        \centering
        \includegraphics[width=0.80\textwidth]{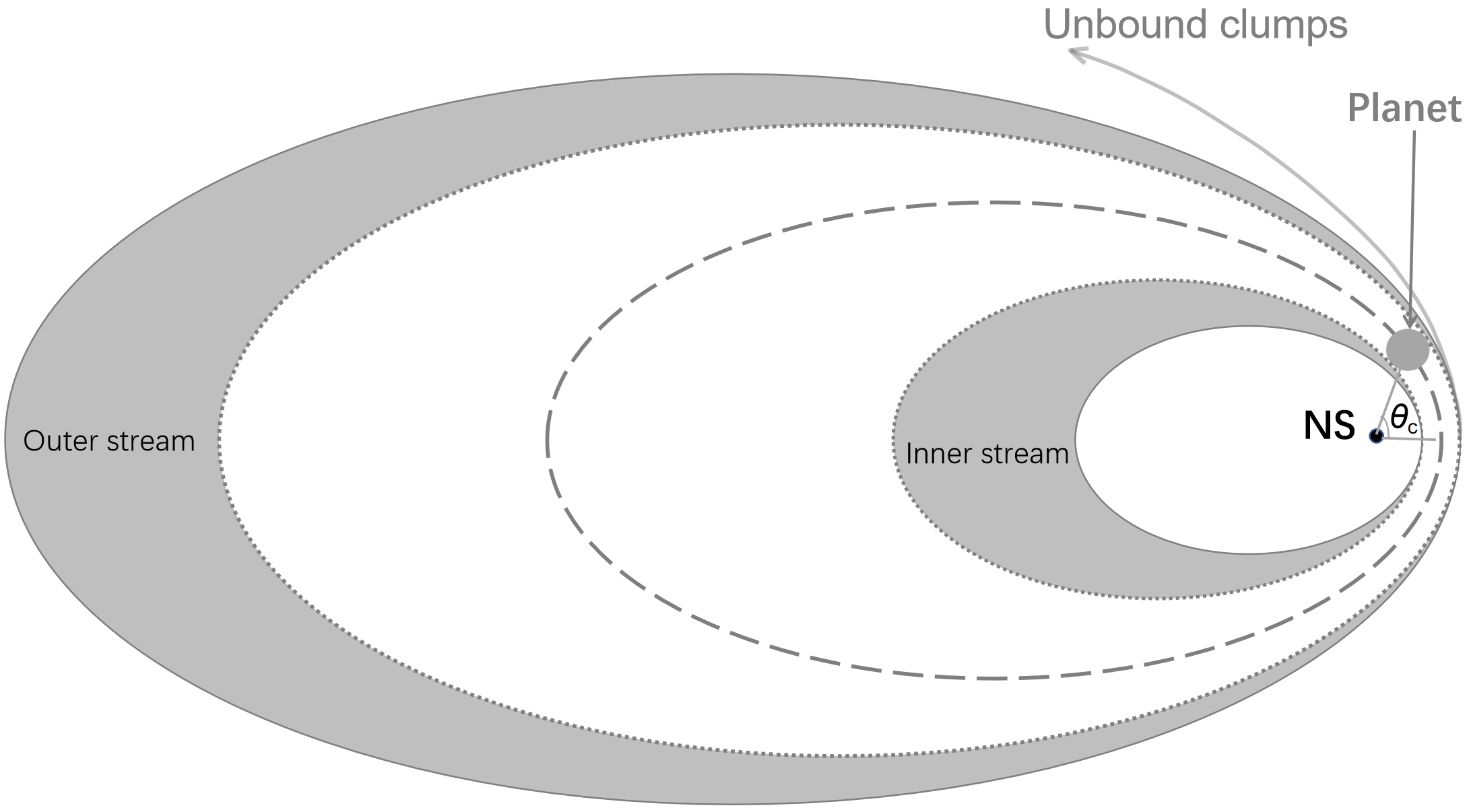}
        \caption{Schematic illustration (not to scale) of the 
          partial disruption of a planet around a NS. The 
          dashed ellipse represents the orbit of the planet. The gray areas 
          represent possible ranges of the clumps' orbit in the inner and outer 
          bound stream. The dotted ellipses at the fringe of these gray 
          areas represent the orbits of the clumps generated 
          near $\theta = \theta_{\rm c}$. The solid ellipse in the inner 
          stream represents the clumps produced near $\theta = 0$, while the
          solid ellipse in the outer stream represents the clumps produced 
          at somewhere between $\theta = 0$ and $\theta = \theta_{\rm c}$, 
          depending on the structure of the planet. The curved arrow 
          represents unbound clumps.
                \label{fig:fig1}}
\end{figure*}

We introduce a NS--planet interaction scenario
to explain the periodic X-ray bursts of SGRs.
We considered a planetary system composed of a NS with a mass
of $M_{\star}$ (we take $M_{\star} = 1.4 M_{\sun}$
and $R_{\star} = 10$ km in the following calculations) and a
rocky planet of mass $m_{\rm pl}$ and radius $R_{\rm pl}$. The planet
orbits around the host NS in a highly eccentric orbit described
by $r=a(1-e^2)/(1+e\cos\theta)$, where $a$ is the semimajor axis
of the orbit, $e$ is the eccentricity, and $ \theta $ is the true
anomaly. The periastron of the orbit is $r_{\rm p}=a(1-e)$.
According to Kepler's third law, the orbital period is related to $a$
as $P_{\rm orb} = \sqrt{ 4 \pi^2 a^{3}/[G(M_{\star}+m_{\rm pl})]}$.
A planet will be tidally disrupted if the tidal force exceeds the
self-gravity during the orbiting process. A planet whose periastron
is only slightly larger than the tidal disruption radius would be partially
disrupted by the tidal force every time it passes through the periastron.
Figure \ref{fig:fig1} presents a schematic illustration of the partial
disruption process.

When the planet passes through the pericenter,
a number of clumps will be generated during the partial disruption process.
The host NS, the surviving main part of the planet, and the clumps
form a many-body system. Ignoring the interactions between the clumps, it
can be simplified as a triple system composed of the NS, the remnant planet,
and a particular clump. In this case, the gravitational perturbation
from the planet plays an important role in the evolution of the clump's orbit,
which could lose its angular momentum due to the perturbation and fall
toward the NS \citep{Kurban2023MNRAS}. The Alfv\'{e}n wave drag caused by
the magnetic field of the NS can further facilitate the in-falling
process \citep[e.g.,][]{Geng2020}. The interaction between the major
clumps and the NS will finally produce a series of X-ray bursts.
We estimate the burst energy, duration, periodicity,
and activity window in Section \ref{sec:xrb}.

\subsection{Key parameters}\label{subsec:para}

In our framework, we considered a rocky planet that is a pure
Fe object, a pure MgSiO$_{3}$ object, or a two-layer planet with
an Fe core and an MgSiO$_{3}$ mantle. The internal structure of such
planets can be determined by using the equations of state (EOSs) that
are widely used in exoplanet modeling (see \citealt{Smith2018NatAs}
for the EOS of Fe materials and \citealt{Seager2007ApJ} for that
of MgSiO$_{3}$ materials). Table \ref{tab:table1} lists some typical parameters
of the three kinds of planets (see \citealt{Kurban2023MNRAS} for more details).

When the size of a planet is larger than 1000 km, 
self-gravity will dominate over the internal tensile stresses \citep{Brown2017MNRAS}.
The tidal disruption radius of such a self-gravity-dominated
planet is $r_{\rm td} = R_{\rm pl} (2M_{\star}/m_{\rm pl})^{1/3}$. 
However, the degree of disruption depends on the impact parameter
($\beta = r_{\rm td}/r_{\rm p}$) as well as the planet's structure.
A slight decrease in the pericenter may result in a complete disruption,
while a small increase will lead to no mass loss at all \citep[e.g.,][]{Guillochon2013ApJ}. 
To be more specific, a planet would be completely destroyed only when the tidal 
force from the NS exceeds the maximum self-gravitational force inside
the planet but not the surface gravitational field. This means that
the dense planet core is most difficult to destroy and could 
survive for slightly larger values of 
$\beta$ \citep[e.g.,][]{Coughlin2022MNRAS,Bandopadhyay2024ApJ}.
In short, a higher central density of the planet makes 
a complete disruption less likely. As a result, partial
disruption could occur in a considerable range of $\beta$ for a planet.
In fact, numerical simulations
\citep{Guillochon2011ApJ, Guillochon2013ApJ, Liu2013ApJ, Nixon2021ApJ}
and analytic derivations \citep{Coughlin2022MNRAS} show that a
partial disruption occurs when 
$ r_{\rm p} \approx 2r_{\rm td}$ ($\beta \approx 0.5$).
In this study, we took this distance as a typical
condition that the planet would be partially disrupted
near the periastron \citep{Kurban2022ApJ, Kurban2023MNRAS}.

\begin{table*}[thb]
        \centering
        \tabcolsep=2.0mm
        \small
        \caption{Typical parameters of the planets considered in our study
                and the timing parameters of the clumps originated from the
                inner side of the planet. }
        \label{tab:table1}
        \begin{tabular}{lcclcccccclccccc}
                \hline\hline
                \noalign{\smallskip}
                \multicolumn{3}{c}{} &\multicolumn{6}{c}{$ P_{\rm orb} = 238 $ days}& &\multicolumn{6}{c}{$ P_{\rm orb} = 398 $ days} \\
                \noalign{\smallskip}
                \cline{4-9}
                \cline{11-16}
                \noalign{\smallskip}
                $ m_{\rm pl} $&
                $ R_{\rm pl} $ &
                $\frac{m_{\rm pl}}{R_{\rm pl}} $&
                $ P_{\rm orb}^{\rm cl} $ &
                $ t_{\rm trav} $&
                $ P_{\rm orb,c}^{\rm cl} $&
                $ t_{\rm trav,c} $&
                $ \Delta P_{\rm orb,up}^{\rm cl} $ &
                $ \Delta t_{\rm trav,up} $&
                &
                $ P_{\rm orb}^{\rm cl} $ &
                $ t_{\rm trav} $&
                $ P_{\rm orb,c}^{\rm cl} $&
                $ t_{\rm trav,c} $&
                $ \Delta P_{\rm orb,up}^{\rm cl} $ &
                $ \Delta t_{\rm trav,up} $
                \\
                
                ($M_{\oplus}$) & 
                ($R_{\oplus}$) & 
                &
                (days) &
                (days) &
                (days) &
                (days) &
                (days) &
                (days) &
                &
                (days) &
                (days) &
                (days) &
                (days) &
                (days) &
                (days)\\
                \hline
                \noalign{\smallskip}
                \multicolumn{16}{c}{Fe planet}\\
                \noalign{\smallskip}
                \cline{1-16}
                \noalign{\smallskip}
                4.6   & 1.2  & 3.8  & 36.3  & 73.1  & 115.1  & 230.3  & 78.8  & 157.2 &  & 41.3  & 83.1  & 154.7  & 309.4  & 113.3  & 226.3  \\
                10.1  & 1.4  & 7.2  & 24.6  & 49.6  & 92.4  & 184.9  & 67.8  & 135.3 &  & 27.2  & 54.7  & 118.9  & 237.8  & 91.6  & 183.1  \\
                18.1  & 1.6  & 11.1  & 17.9  & 35.9  & 75.6  & 151.3  & 57.8  & 115.4 &  & 19.3  & 38.8  & 94.0  & 188.0  & 74.7  & 149.2  \\                
                \cline{1-16}
                \noalign{\smallskip}
                \multicolumn{16}{c}{MgSiO$_{3}$ planet}\\
                \noalign{\smallskip}
                \cline{1-16}
                \noalign{\smallskip}
                4.4   & 1.7  & 2.6  & 54.9  & 111.1  & 141.7  & 283.8  & 86.8  & 172.8 &  & 65.3  & 131.6  & 200.1  & 400.5  & 134.8  & 268.8  \\
                9.9   & 2.1  & 4.7  & 40.1  & 80.8  & 121.4  & 242.9  & 81.3  & 162.0 & & 46.0  & 92.7  & 165.0  & 330.1  & 119.0  & 237.4  \\
                18.2  & 2.5  & 7.3  & 30.8  & 62.1  & 105.3  & 210.8  & 74.5  & 148.7 & & 34.6  & 69.6  & 138.9  & 277.9  & 104.3  & 208.3  \\
                \cline{1-16}
                \noalign{\smallskip}
                \multicolumn{16}{c}{Two-layer planet}\\
                \noalign{\smallskip}
                \cline{1-16}
                \noalign{\smallskip}
                3.3   & 1.4  & 2.4  & 53.8  & 108.7  & 140.4  & 281.0  & 86.5  & 172.3 & & 63.8  & 128.6  & 197.7  & 395.5  & 133.8  & 267.0  \\
                9.8   & 1.8  & 5.4  & 34.4  & 69.4  & 112.0  & 224.0  & 77.5  & 154.6 & & 39.0  & 78.5  & 149.6  & 299.2  & 110.5  & 220.7  \\
                12.8  & 1.6  & 8.0  & 24.0  & 48.3  & 91.0  & 181.9  & 67.0  & 133.7 & & 26.4  & 53.1  & 116.6  & 233.3  & 90.2  & 180.2  \\
                17.1  & 2.4  & 7.1  & 31.1  & 62.6  & 105.8  & 211.7  & 74.7  & 149.1 & & 34.9  & 70.2  & 139.6  & 279.3  & 104.8  & 209.2  \\
                \hline
                \noalign{\smallskip}
        \end{tabular}
        \begin{minipage}{0.98\textwidth}
                {Note: The timing parameters of the clumps are calculated for two cases of the planet's orbital period
                        ($ P_{\rm orb} = 238, 398 $ days) according to the methods given in the main text.}        
        \end{minipage}
\end{table*}

Taking $ r_{\rm p} \sim 10^{11}$ cm, the 
eccentricity is $e \sim 0.992$ for a planet with an
orbital period of $P_{\rm orb} = 240$ day ($a = 0.85$ au), while it is
$e \sim 0.994$ for $P_{\rm orb} = 400$ day ($a = 1.19$ au).
One may obtain a slightly wider parameter range when the diversity 
of the planet structure is considered \citep{Kurban2022ApJ,Kurban2023MNRAS}. 
Such a highly eccentric orbit satisfying  
$ r_{\rm p} \approx 2r_{\rm td}$ can be formed in the dynamical processes
mentioned earlier. A simulation by \cite{Goulinski2018MNRAS}
showed that more than 99.1\% of the captured planets form orbits
with $ 0.85 < e \lesssim1 $ and $a \sim 1$ --- $10^4$ au. 
The planet-planet scattering \citep[e.g.,][]{Carrera2019AA} and the
capture of a planet from a nearby planetary system via the Hills mechanism
\citep[e.g.,][]{Cufari2022ApJ} are the other two efficient ways to form highly
eccentric orbits. In addition, an extremely high eccentricity can also be reached due
to the Kozai-Lidov mechanism for a wide range of semimajor axis
\citep[e.g.,][]{Naoz2016ARAA}.

The clumps generated during the disruption will have slightly
different orbits (see Figure \ref{fig:fig1}), which are determined by the planet's
binding energy and pericenter distance \citep{Norman2021ApJ}. 
For example, the clumps originating from the inner side
of the planet will have a relatively small orbit.
The semimajor axis of their orbits can be
expressed as \citep{Malamud2020a,Brouwers2022MNRAS}
\begin{equation}\label{a_prime}
a_{\rm cl} = a \left( 1 + a\frac{2R}{d(d - R)}\right)^{-1}~~(\rm for~inner~orbit~clumps)
,\end{equation}
where $ a $ is the planet's original semimajor axis, $ d $ is
the distance between the NS and the planet at the moment of the breakup
(here $ d = r_{\rm p}$), $ R $ is the displacement of the clump relative
to the planet's mass center at the moment of breakup ($ R = 0 $ corresponds
to the center of the planet). 
Because the planet is in a bound orbit, the clumps in the inner
stream are generally bound to the NS so that they can be 
relevant to the X-ray bursts studied in the work.

For clumps in the outer orbit stream,
their semimajor axes can be calculated as 
$a_{\rm cl} = a \left( 1 - a\frac{2R}{d(d + R)}\right)^{-1}$.
From this relation, a critical displacement can be derived as 
$ R_{\rm crit} = d^{2}/(2a - d)$. 
A clump is still bound to the NS if its displacement is  $ R < R_{\rm crit} $,
while it is unbound for $ R > R_{\rm crit} $ \citep{Malamud2020a}.
$ R_{\rm crit}$ depends on the pericenter and binding energy of the planet. 
Combining the partial disruption condition of $d = r_{\rm p} = 2r_{\rm td}$,
we can assess the boundness of the clumps in the outer stream. 
It is found that unbound (free) clumps could be generated in the 
following cases considered here: an Fe planet with $ m_{\rm pl} = 18.1\,M_{\oplus}$
in the case of $P_{\rm orb} = 238$ day; an Fe planet with $ m_{\rm pl} = 10.1,18.1\,M_{\oplus}$,
an MgSO$_{3}$ planet with $ m_{\rm pl} = {18.2}\,M_{\oplus}$, 
and a two-layer planet with $ m_{\rm pl} = {12.8}\,M_{\oplus}$
for the case $P_{\rm orb} = 398$ day.
The outer stream clumps generated from other planets considered in our study are all
bound to the NS. The orbits of these bound clumps will also be
affected by the remnant planet's gravity and can be altered
significantly due to the scattering effect. They may even be ejected
from the system finally. As a result, the clumps in the outer 
stream are less likely to collide with the NS (especially on short 
timescales), and their involvement 
in the generation of X-ray bursts is not expected.

In the case of a full disruption, the material in a 
bound orbit returns to the periastron on a timescale of 
$T_{\rm ret} = (r_{\rm td}^{2}/2R)^{3/2}(2\pi/\sqrt{GM_{\star}})$
\citep{Lacy1982ApJ,Norman2021ApJ}. 
We note that 
this expression underestimates the return time if $r_{\rm td}$ is 
replaced by $r_{\rm p}$ for deep 
encounters \citep{Rossi2021SSRv,Coughlin2022MNRAS}. For the 
partial disruption in our framework, the clump's return time  
approximately equals its orbital period, 
$ t_{\rm ret} \approx P_{\rm orb}^{\rm cl} 
= \sqrt{4 \pi^2 a_{\rm cl}^3/[G (M_{\star}
        + m_{\rm cl})]}$. The clump's orbit should satisfy
$ r_{\rm p} = a_{\rm cl}(1-e_{\rm cl}) \pm R$
(here ``+'' and ``-'' for the clumps in the inner and outer stream, respectively),
from which one can derive the eccentricity as 
$ e_{\rm cl} = 1 - (r_{\rm p} \mp R)/a_{\rm cl} $.
Since the clumps are stripped off from the surface of the planet, we
took $ R \sim R_{\rm pl}$ when calculating their orbital parameters.

After the partial disruption, clumps of different sizes 
may be generated. Larger homogeneous monolithic clumps generally
have larger cohesive strength \citep[e.g.,][]{Malamud2020a}. Smaller 
ones may merge to form larger clumps due to gravitational instabilities
in the hydrodynamical evolution of the debris disk \citep{Coughlin2015ApJ}, 
which itself depends on the EOS and is preferred for EOSs stiffer 
than $\gamma = 5/3$ \citep{Coughlin2016MNRAS_a,Coughlin2016MNRAS_b,
        Coughlin2020ApJ,Coughlin2023MNRAS,Fancher2023MNRAS}. 
In our cases, the characteristics of the massive clumps formed through the
accumulation of debris due to gravitational instabilities 
are similar to that of the rubble pile asteroids, which have a
material strength in the range of $\sim$ 1 Pa --- 1000 Pa \citep{Walsh2018ARAA}.

\cite{Veras2014MNRAS} have investigated the tidal disruption of 
rubble pile asteroids by assuming that the asteroid is frictionless,
which means they essentially ignored the cohesive strength. The classical 
Roche approximation based solely on self-gravity could be applied in that 
case. However, they have also pointed out that cohesive strength could play a 
role in the process. Especially, the breakup distance of small bodies ($<1000$ km) is
mainly determined by the material strength \citep{Brown2017MNRAS}.
The larger the cohesive strength is, the closer it gets to the central 
star \citep{Brouwers2022MNRAS}. The homogeneous monolithic clumps and
the rubble pile clumps formed in our cases have different material
strengths so that they break up at different distances that are smaller than
the partial disruption radius of the planet. In short, the clumps can
remain intact even when they are much closer to the NS due to their
cohesive strength. For them, the breakup distance is \citep{Zhang2021ApJ91591Z}
\begin{equation} \label{eq:td_for_largeC}
        r_{\rm str} = \left( \frac{\sqrt{3} GM_{\star} r_{\rm cl}^2 \rho_{\rm cl}}{5k}\right)^{1/3},
\end{equation}
where $ r_{\rm cl} $ and $ \rho_{\rm cl} $ are the radius
(here we assumed that the clump is spherical for the sake of simplicity)
and density of the clump, respectively; $k$ is a function
of the internal friction angle, $\phi$, and the cohesive strength,
$ C $: $ k = 6C \cos\phi/\sqrt{3}(3 -\sin\phi)$.
The friction angle of geological materials commonly ranges
from $25^{\circ}$ to $50^{\circ}$
\citep[e.g.,][]{Holsapple2008Icar,Bareither2008,Jiang2018,Villeneuve_Heap_2021}.
The typical strength is $C \sim 1$ Pa for comets
\citep{Gundlach2016AA} and $C \sim 1$ Pa --- 1000 Pa for rubble
pile asteroids \citep{Walsh2018ARAA}, while it is in the range 0.1 Mpa --- 50 MPa for monolithic asteroids/meteorites
\citep{Martin2009, Ostrowski2019, Pohl2020MPS, Veras2020MNRAS, Villeneuve_Heap_2021}.
As a result, the breakup distance for the clumps considered here is
$\sim 10^9$ --- $10^{10}$ cm.

\section{X-ray bursts}\label{sec:xrb}

\subsection{Periodicity and activity window}\label{subsec:period-active}

\begin{figure}[!h]
        \centering
        \includegraphics[width=0.90\columnwidth]{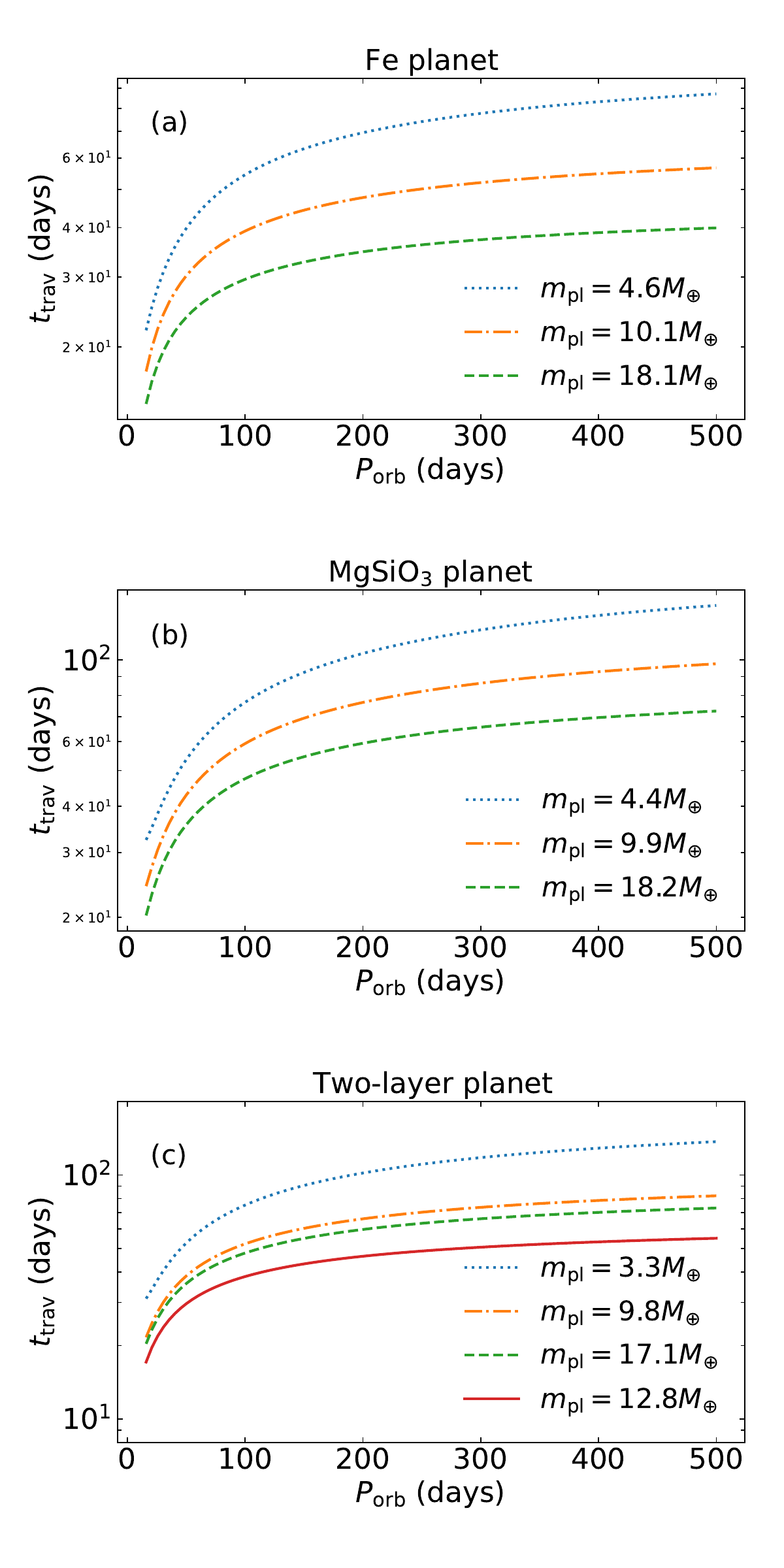}
        \caption{Travel time ($t_{\rm trav}$) of the clumps in the
                innermost orbit as a function of the planet's orbital period ($P_{\rm orb}$).
                Three kinds of planets are considered: (a) Fe planets, (b) MgSiO$_{3}$ planets, and
                (c) two-layer planets. The mass of the planet ($m_{\rm pl}$) is
                marked in each panel.
                \label{fig:fig2}}
\end{figure}

The periodicity and active window of X-ray bursts are determined
by the orbital evolution of the clumps generated in the NS planet
system. It can be simplified as a triple system composed
of the NS, the remnant planet, and a particular clump.
The clump's orbit evolves under the influence of strong gravitational
perturbation from the remnant planet.

In a triple system where a test particle revolves
around its host in a close inner orbit while a third object
moves around in an outer orbit, the eccentricity of
the test particle can be significantly altered by the
outer object. 
This is known as the Kozai-Lidov mechanism, which
can explain the long-term orbital evolution of  
triple systems in the hierarchical limit \citep{Kozai1962,Lidov1962}. 
In recent decades, it has been further developed and applied to 
various mildly hierarchical or non-hierarchical systems.
The hierarchy of a system is mainly measured by a parameter 
defined as $ \epsilon = a_{1} e_{2}/[a_{2}(1 - e_{2}^2)]$, which 
actually is the coefficient of the octupole-order interaction term
\citep{Lithwick2011ApJ,Li2014ApJ}. 
Here the eccentricity ($e$) and semimajor axis ($a$) of the inner
and outer orbits are denoted by subscripts 1 and 2, respectively. 
$ \epsilon$ parameterizes the size of the external orbit versus 
that of the internal orbit. The initial orbital parameters of
a triple system have an obvious influence on its final fate after a 
long-term dynamic evolution. A system with $\epsilon \leq 0.1 $ is 
hierarchical and stable, a system with $\epsilon > 0.3 $ is 
nonhierarchical and unstable, while $0.1 <\epsilon \leq 0.3 $ 
corresponds to the mildly hierarchical condition \citep{Naoz2016ARAA}. 
We note that the Kozai-Lidov mechanism (secular interaction) is broken down for
the nonhierarchical and unstable systems so that it cannot provide a
meaningful description for the orbital evolution at all \citep{Perets2012ApJ,Katz2012}.   
Strong perturbations from the outer orbit object \citep{Toonen2022AA}
lead the test particle to effectively lose its angular momentum on a
timescale on the order of the inner orbit period, causing it to collide 
with the central object \citep[e.g.,][]{Antonini2014ApJ,Antonini2016ApJ,He2018MNRAS,Hamers2022ApJ}.

In our framework, the parameters of the clumps in the 
inner stream satisfy $\epsilon > 0.3 $, which means that they 
are in a nonhierarchical and unstable regime. The secular 
approximation (Kozai-Lidov mechanism) is broken down for them. 
The surviving major portion of the planet plays the role of
the outer object, which can significantly alter the
clump's orbit and cause it to fall onto the NS on a timescale
comparable to a few orbital periods \citep{Kurban2023MNRAS}.
In fact, in a recent study, it was shown that energy exchange 
between the clump and the remnant planet could occur due to 
strong gravitational interactions, which leads the clump's 
periastron distance ($r_{\rm p}$) to abruptly change to a 
very small value \citep[e.g.,][]{Zhang2023ApJ}. 
Below, we present a rough estimate for the travel time 
that a clump experiences from its birth to
the final collision with the NS. The total duration for
most clumps to collide with the NS (i.e., the active window) can
also be estimated.

\begin{figure*}
        \centering
        \includegraphics[width=0.80\textwidth]{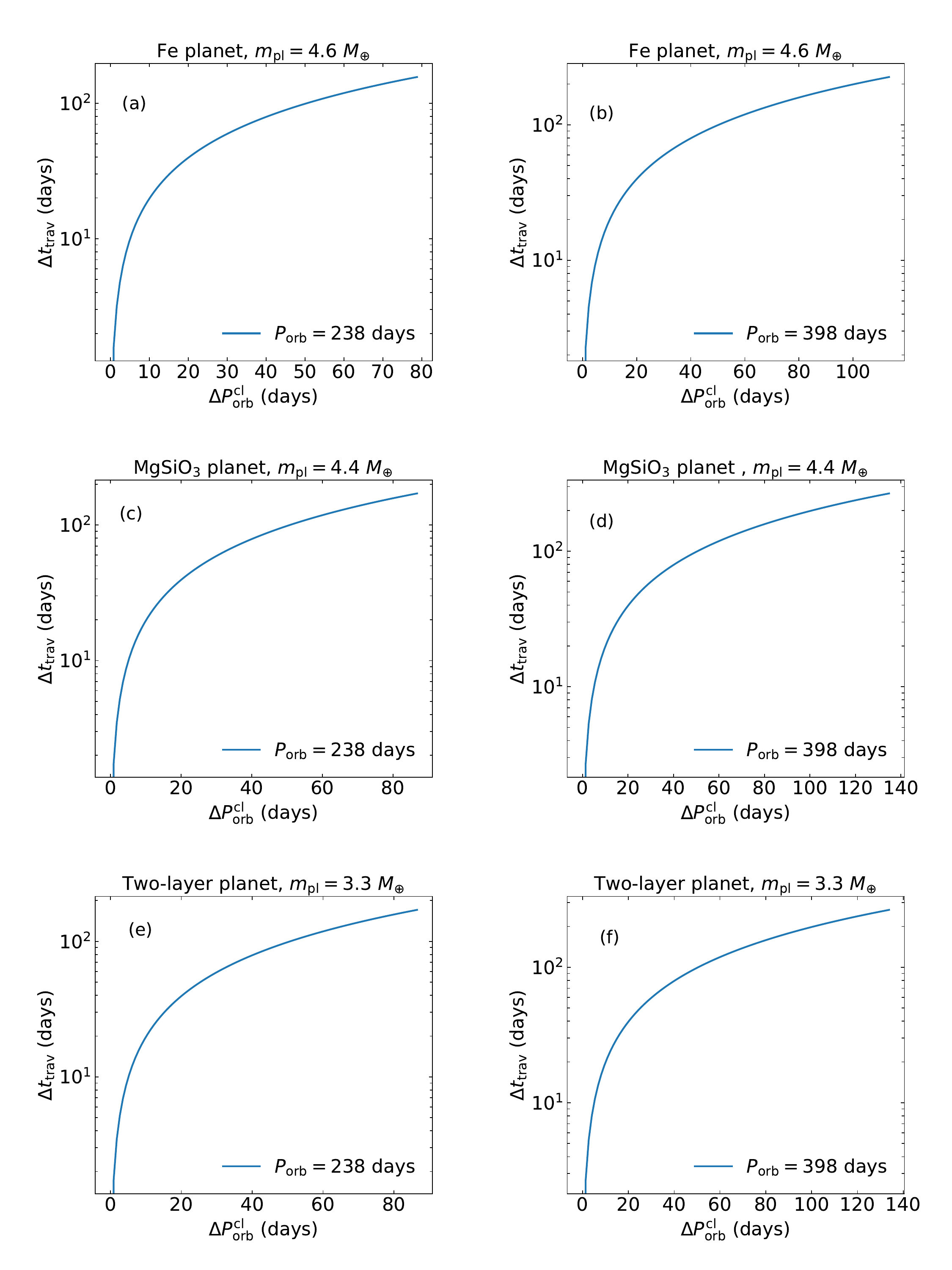}
        \caption{Difference in the travel time ($\Delta t_{\rm trav}$) as a function of the clump's
                orbital period difference ($\Delta P_{\rm orb}^{\rm cl}$) for planets in various conditions.
                The upper panels show the cases of Fe planets for $P_{\rm orb} = 238$ days (Panel a)
                and $P_{\rm orb} = 398$ days (Panel b).
                The middle panels show the cases of MgSiO$_{3}$ planets for $P_{\rm orb} = 238$ days (Panel c)
                and $P_{\rm orb} = 398$ days (Panel d).
                The lower panels show the cases of two-layer planets for $P_{\rm orb} = 238$ days (Panel e)
                and $P_{\rm orb} = 398$ days (Panel f). Note that the endpoint of each
                line corresponds to $\Delta P_{\rm orb,up}^{\rm cl}$ and $\Delta t_{\rm trav,up}$.
                \label{fig:fig3}}
\end{figure*}

After the partial disruption, a clump originating from the inner side of
the planet will move in an inner orbit. The travel time for it to
fall onto the NS depends on its return time ($ t_{\rm ret}$) to the periastron
as well as the evolution timescale of its angular momentum ($t_{\rm evo}$).
The return time approximately equals its orbital period,  
$ t_{\rm ret} \approx P_{\rm orb}^{\rm cl} $.
After the formation of the inner orbit, the clump loses its angular momentum ($ j_{\rm cl} = \sqrt{1 - e_{\rm cl}^2}
\rightarrow  j_{\rm cl} \sim 0 $) on a timescale of $t_ {\rm evo} $ due to
gravitational perturbations. This is expected to occur when the
clump passes through the periastron at the end of the second or third orbit
because $t_ {\rm evo}$ is much less than $P_{\rm orb}^{\rm cl}$ under the 
condition of $m_{\rm pl} > 2 M_{\oplus}$ and $P_{\rm orb} > 100$ day \citep{Kurban2023MNRAS}.
Therefore, the clump's travel time from its birth at the
planet to its collision with the NS can be approximated as \citep{Kurban2023MNRAS}
\begin{equation}\label{eq:ttrav}
        t_{\rm trav} \approx 2 P_{\rm orb}^{\rm cl} + t_{\rm evo},
\end{equation}
where $t_{\rm evo}$ is expressed as \citep[e.g.,][]{Antonini2014ApJ}
\begin{equation}\label{eq:evo}
        t_{\rm evo} = \left(\frac{1}{j_{\rm cl}}\frac{dj_{\rm cl}}{dt}\right)^{-1} \approx
        P_{\rm orb}^{\rm cl}\frac{1}{5\pi} \frac{M_{\star}}{m_{\rm pl}}
        \left[ \frac{a(1 - e)}{a_{\rm cl}} \right]^3 \sqrt{1 - e_{\rm cl}}.
\end{equation}
We note that different clumps with slightly different orbital periods
will arrive at the NS at different times.

Figure \ref{fig:fig2} shows the travel time of a clump in the
innermost orbit for pure Fe planets, MgSiO$_{3}$ planets, and two-layer planets.
From this figure, we see that the clumps will fall onto
the NS on a short timescale on the order of its own orbital period,
which is affected by the orbit and composition of
the planet.

The activity widow of the bursts can be estimated by considering the
dispersion of the arrival time of various clumps with slightly different
orbital periods. We can consider the case of two clumps:
one has an orbital period of $P_{\rm orb}^{\rm cl}$, the other
has a different orbital period of $P_{\rm orb}^{\rm cl} + \Delta P_{\rm orb}^{\rm cl}$,
where $\Delta P_{\rm orb}^{\rm cl} $ is their difference in orbit.
Their difference in arrival time can be calculated from the derivative of
travel time as
$\Delta t_{\rm trav} \approx 2 \Delta P_{\rm orb}^{\rm cl} -
4t_{\rm evo}(\Delta P_{\rm orb}^{\rm cl}/P_{\rm orb}^{\rm cl})/3$,
which is just the time interval between two successive collisions.

The simulations by \citet{Malamud2020a, Malamud2020b} show that the
distribution of the orbital period of inner bound clumps can range from
$P_{\rm orb}^{\rm cl}$ to $P_{\rm orb}$ in a full disruption. However, the
competition between the tidal force and the self-gravity continuously
evolves over the encounter. The largest partial disruption distance for
a planet is $\sim 2.7r_{\rm td}$ \citep{Guillochon2011ApJ,Liu2013ApJ}.
During a strong encounter, the mass loss from the planet continues
till a separation of several times $r_{\rm td}$ \citep[e.g.,][]{Ryu2020ApJ}.
For the partial disruption cases discussed here, 
the mass loss will continue until the planet's separation
from the NS is $r_{\rm c} \sim 4r_{\rm td}$. In other words, 
the mass loss occurs until the phase $ \theta \sim \theta_{\rm c}$, 
where $\theta_{\rm c}$ is the true anomaly at $r = r_{\rm c}$.
When $\theta > \theta_{\rm c}$ (or $r > r_{\rm c}$), new
clumps could not be generated till the next periastron passage.
For the inner and outer bound streams, the clumps marginally
bound to the remnant planet will be re-accreted by the surviving core
of the planet after a few dynamical timescales defined by
$t_{\rm dyn} = \sqrt{R_{\rm pl}^{3}/Gm_{\rm pl}}$ \citep[e.g.,][]{Guillochon2011ApJ,Coughlin2019ApJ}. 
We note that the dynamical timescale is 
typically $t_{\rm dyn} \sim $ 6 --- 14 minutes, which depends on the structure of
the planets. On the other hand, the clumps that are not bound to the 
remnant planet will return to the periastron and continue to orbit 
around the NS \citep[e.g.,][]{Guillochon2011ApJ}.
One can obtain the critical semimajor axis
$a_{\rm cl,c}$ (or the critical orbital period $P_{\rm orb,c}^{\rm cl}$)
using the critical separation of $d = r_{\rm c} \sim 4r_{\rm td}$.
Thus, for the clumps in the inner orbits, we expect their orbital periods
to be between $P_{\rm orb}^{\rm cl}$ and $P_{\rm orb,c}^{\rm cl}$.
The upper limit of $\Delta P_{\rm orb}^{\rm cl}$ would then be
$\Delta P_{\rm orb,up}^{\rm cl} = P_{\rm orb,c}^{\rm cl} - P_{\rm orb}^{\rm cl} $.
We used this parameter to estimate $ \Delta t_{\rm trav,up} $, which is
just the activity window of X-ray bursts in our framework.

Figure \ref{fig:fig3} illustrates $\Delta t_{\rm trav}$ as a function
of $\Delta P_{\rm orb}^{\rm cl}$ for planets with different orbital
period and composition. We note that $\Delta P_{\rm orb}^{\rm cl}$ satisfies
$0 < \Delta P_{\rm orb}^{\rm cl} \lesssim \Delta P_{\rm orb,up}^{\rm cl}$,
which determines the range of the active window for the X-ray bursts.
$\Delta t_{\rm trav,up}$ is sensitive to both
$r_{\rm p}$ and $r_{\rm c}$. A too-small $r_{\rm p} $ may cause the
period to completely disappear.

Table \ref{tab:table1} lists the timing parameters of the clumps disrupted from
various planets with different compositions, masses, and orbital periods.
$P_{\rm orb}^{\rm cl}$ and $P_{\rm orb,c}^{\rm cl}$ are the orbital periods
of the clumps in the innermost orbit and the clumps stripped off from the
planet at $ \theta_{\rm c}$ (or $r=r_{\rm c}$), respectively.
$t_{\rm trav}$ and $t_{\rm trav,c}$ are their travel times, and
$\Delta P_{\rm orb,up}^{\rm cl}$ and $ \Delta t_{\rm trav,up} $ are
their orbital period difference and travel time difference, respectively.
It can be seen from this table that these parameters are affected
by the structure and orbital period of the planet. Generally, a larger
$P_{\rm orb}$ leads to a larger $\Delta t_{\rm trav,up}$. For planets with
a particular $P_{\rm orb}$, the structure is also an important factor:
$\Delta t_{\rm trav,up}$ decreases with the increase in compactness
($m_{\rm pl}/R_{\rm pl}$).

\subsection{Energy and duration}\label{subsec:E-t}

In the last section we investigated the properties of clumps and
show that the clumps fall onto the NS on a short timescale.
The collision of a small body (asteroid) with a NS can produce X-ray bursts
\citep{Huang2014, Geng2020, Dai2016, Dai2020b}.
The gravitational potential energy of the small body will
transform to X-ray burst energy as \citep[e.g.,][]{Geng2020,Dai2020b}
\begin{equation}\label{eq:EX}
        E_{X} = \eta\frac{GM_{\star}m_{\rm cl}}{R_{\star}},
\end{equation}
where $\eta $ ($\eta < 1$) is the energy transforming efficiency.

In our case, the range of the mass of the clumps can be relatively wide.
For example, simulations show that the size of clumps generated during
the partial disruption ranges from a few kilometers to $\sim 100$
km \citep{Malamud2020a, Malamud2020b}. It depends on the distance to
the NS as well as its intrinsic material strength. On the other hand,
if a clump is too large, it cannot resist the tidal force and will further
break up during its falling toward the NS. The clumps will interact with
the magnetosphere and produce Alfv\'{e}n wings at the light cylinder radius,
$R_{\rm LC} = cP_{\rm spin}/2\pi$ 
\citep{Cordes2008ApJ, Mottez2011AA, Mottez2013AAa,Mottez2013AAb,Chen2022ApJ},
which further helps the NS capture the
clumps \citep{Geng2020} at the Alfv\'{e}n radius.
The Alfv\'{e}n radius is determined by assuming that the kinetic
energy equals the magnetic energy, namely 
$\rho_{\rm cl}\upsilon_{\rm R_{A}}^{2}/2 = B_{\rm R_{\rm A}}^2/8\pi$,
where $\upsilon_{\rm R_{A}} = \sqrt{2GM_{\star}/R_{\rm A}}$ and
$B_{\rm R_{\rm A}} = B_{\star}(R_{\star}/R_{\rm A})^3$ are the free-fall velocity
and magnetic field strength at the distance $R_{\rm A}$ from the NS, respectively.
The Alfv\'{e}n radius $R_{\rm A}$ then can be calculated as
$R_{\rm A} = (B_{\star}^2 R_{\star}^6/8\pi GM_{\star}\rho_{\rm cl})^{1/5}$. Taking
$R_{\star}$ = 10 km and $\rho_{\rm cl} =$ 5 --- 8 g cm$^{-3}$ (the surface density of
the planet with different compositions), we get $R_{\rm A} = $ (4.43 --- 4.04) $\times10^{7}$ cm for SGR 1935+2154
and $R_{\rm A} = $ (1.83 --- 1.67) $\times10^{7}$ cm for SGR 1806-20.

\begin{figure}
        \centering
        \includegraphics[width=0.95\columnwidth]{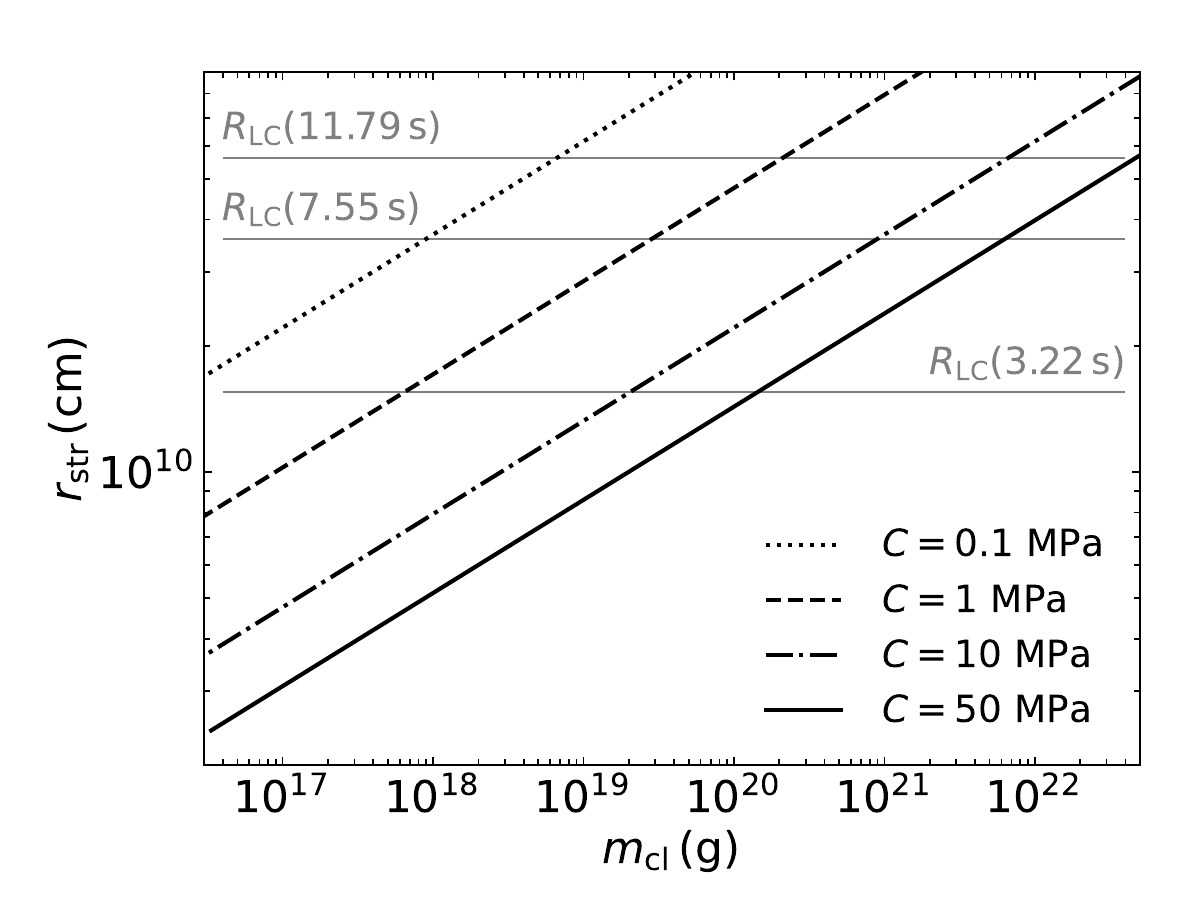}
        \caption{Breakup distance as a function of the clump mass. The clump is a rocky (MgSO$_{3}$) object
                with a density of $\rho_{\rm cl} =5\rm\,g\,cm^{-3}$ and an internal friction angle of $ \phi = 45^{\circ} $.
                The dotted, dashed, and dash-dotted line corresponds to the cohesive strength of $ C =$ 0.1, 1, and 10 MPa,
                respectively. The solid line represents an iron clump with $\rho_{\rm cl} = 8\rm\,g\,cm^{-3}$,
                $ C = 50 $ MPa, and $\phi = 37^{\circ}$.
                The horizontal lines represent the light cylinder radii for NSs with different spin periods.
                \label{fig:fig4}}
\end{figure}

Depending on the size and mass of the clumps near $R_{\rm LC}$,
X-ray bursts with different energies can be produced. Following
Equation (\ref{eq:td_for_largeC}), we can calculate the upper limit of the
clump size at the light cylinder. Figure \ref{fig:fig4} shows the breakup
distance as a function of the clump mass. The dotted, dashed, and dash-dotted
lines correspond to MgSO$_{3}$ clumps ($\rho_{\rm cl} = 5\rm\,g\,cm^{-3}$) with
cohesive strengths of $C = 1 $ kPa, 1 MPa, and 10 MPa, respectively.
These $C$ values are typical for rocky materials
\citep{Gundlach2016AA, Pohl2020MPS, Veras2020MNRAS, Villeneuve_Heap_2021}.
Here, an internal friction angle of $\phi = 45^{\circ}$ is used for rocky materials
\citep[e.g.,][]{Holsapple2008Icar,Bareither2008,Villeneuve_Heap_2021}.
The solid line represents iron clumps ($\rho_{\rm cl} = 8\rm\,g\,cm^{-3}$),
for which $C = 50 $ MPa \citep{Ostrowski2019, Pohl2020MPS}
and $\phi = 37^{\circ}$ \citep{Jiang2018} are adopted.
The horizontal lines represent the light cylinder radius
($ R_{\rm LC}$) of NS with a spin period of
11.79 s (1E 1841-045, \citet{Dib2014ApJ}), 7.55 s (SGR 1806-20),
and 3.245 s (SGR 1935+2154), respectively.
From Fig. \ref{fig:fig4}, we see that the mass range of
the clumps that safely enter $R_{\rm LC}$ is up to $\sim 10^{22} $ g,
which satisfies the energy budget of X-ray bursts.
We notice that $R_{\rm LC} \sim 10^{10}$ cm is much larger than $R_{\rm A}$.
But this is not a problem in our case, since the clump can effectively lose
its angular momentum due to perturbation when it is still far from the light
cylinder.

\begin{figure}
        \centering
        \includegraphics[width=0.95\columnwidth]{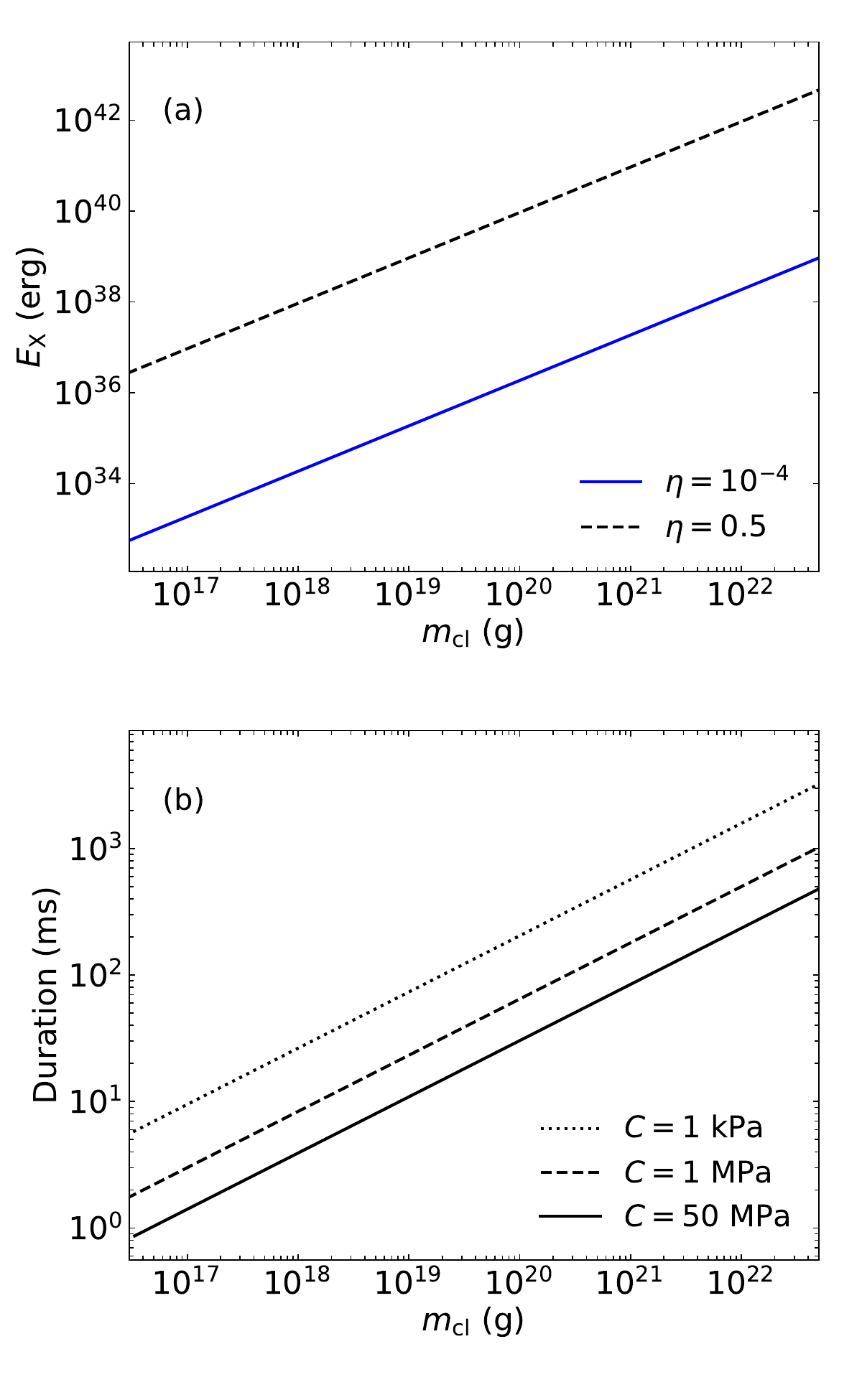}
        \caption{Energy (Panel a) and duration (Panel b) of X-ray bursts as a function of
                the clump mass. In Panel (a) the dashed and solid lines correspond to an energy
                transforming efficiency of $ \eta = 0.5 $ and $ \eta = 10^{-4}$, respectively.
                In Panel (b) the dotted and dashed lines represent the MgSO$_{3}$ clumps
                with a cohesive strength of $ C = 1$ kPa and $ C = 1$ MPa, respectively.
                The solid line corresponds to iron clumps with
                $ C = 50$ MPa. \label{fig:EX}}
\end{figure}

The energies of the X-ray bursts produced through the interactions between
the clumps and the NS can be calculated from Equation (\ref{eq:EX}).
The results are plotted in Panel (a) of Fig. \ref{fig:EX}. It shows
that the energetics can be up to $\sim10^{43}$ erg, which
is large enough to account for the energies of the short X-ray bursts
observed from SGR 1935+2154 and SGR 1806-20.

The burst duration can be calculated as \citep[e.g.,][]{Dai2020b}
\begin{equation}\label{eq:dt}
        \Delta t = \frac{12r_{\rm cl}}{5}\left( \frac{r_{\rm str}}{GM_{\star}}\right) ^{1/2},
\end{equation}
where $r_{\rm str}$ is the breakup distance of the clumps given by Equation (\ref{eq:td_for_largeC}).
Here, a relation of $r_{\rm cl} = (3m_{\rm cl}/4\pi\rho_{\rm cl})^{1/3}$
is adopted. According to Equation (\ref{eq:dt}), the burst duration
can range from a few milliseconds to
seconds, as shown in Panel (b) of Fig. \ref{fig:EX}.
We see that the burst duration is also compatible with the observations.

\section{Comparison with observations}\label{sec:comparison}
The model described in Section \ref{sec:xrb} can explain the
periodical X-ray bursts observed from SGRs. Here
we confront the theoretical burst energy, duration, period,
and activity window with the observations.

\subsection{SGR 1935+2154}
\label{subsec:SGR1935+2154}

\begin{figure}
        \centering
        \includegraphics[width=0.95\columnwidth]{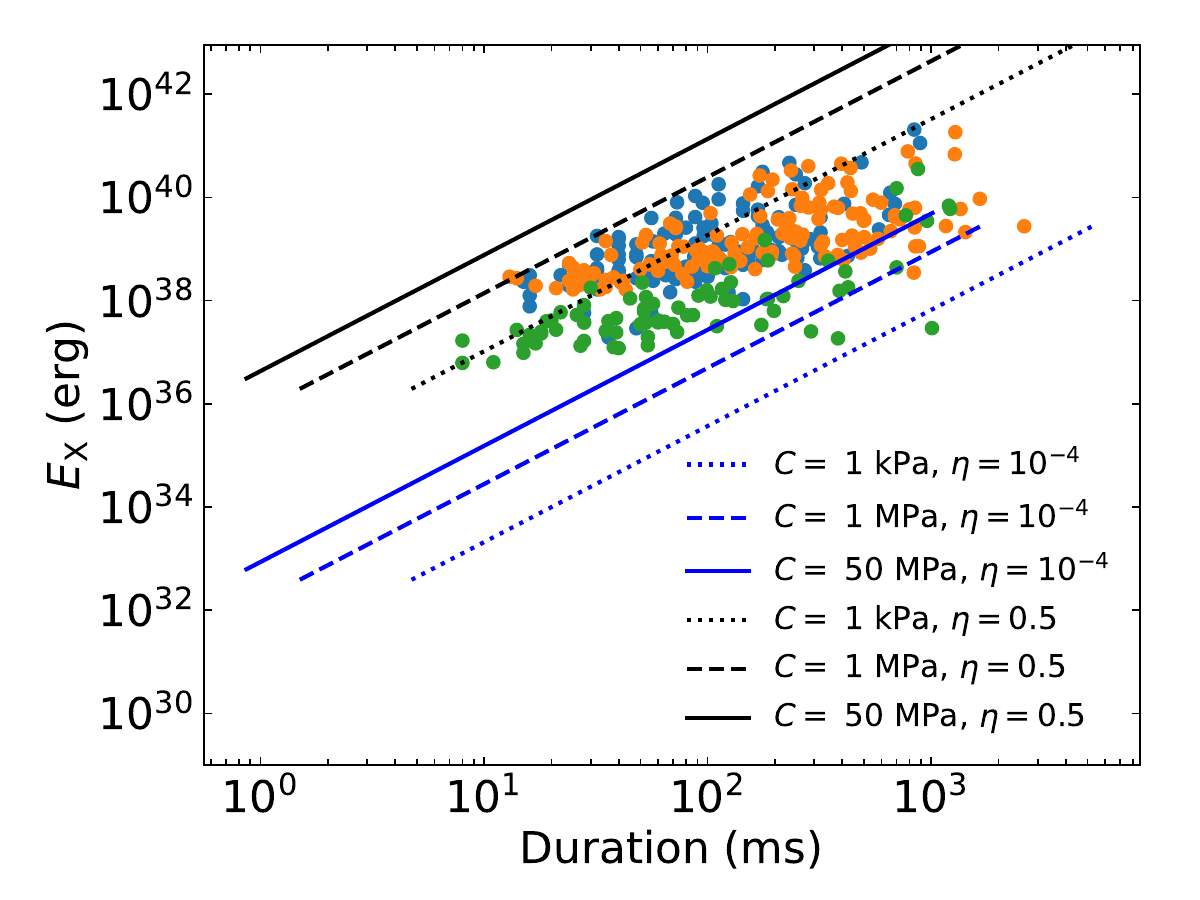}
        \caption{X-ray bursts from SGR 1935+2154 plotted on the energy-duration
                plane. The orange \citep{Lin2020ApJ_b}, blue \citep{Lin2020ApJ_a}, and green
                \citep{Cai2022ApJS_a,Cai2022ApJS_b} dots represent the bursts
                observed by various groups. The lines of different styles are the
                theoretical results calculated by taking different values for the cohesive
                strength ($C$) and energy transforming efficiency ($\eta$).
                \label{fig:1935}}
\end{figure}

Figure \ref{fig:1935} illustrates the burst energy as a function of
duration for SGR 1935+2154. The observational data points of SGR
1935+2154 are shown by assuming a distance of 9 kpc for the source
\citep{Lin2020ApJ_a,Lin2020ApJ_b,Cai2022ApJS_a,Cai2022ApJS_b}.
The solid lines are calculated for different $C$ and $\eta$ values.
We see that the energy budgets can be easily satisfied when the
two parameters are evaluated in reasonable ranges.

The periodicity and activity window of the X-ray bursts from SGR
1935+2154 can also be explained. From Fig. \ref{fig:fig3}, we see
that when we take the orbit period as $P_{\rm orb} = 238$ days and the
upper limit of the period difference
as $\Delta P_{\rm orb,up}^{\rm cl} = \Delta P_{\rm orb}^{\rm cl} \approx 75$ days,
then the arrival time difference is $\Delta t_{\rm trav} = 150$ days. This
agrees well with the observed $\sim$150 day activity window of the SGR.

\subsection{SGR 1806-20}\label{subsec:SGR1806-20}

More than 3000 X-ray bursts from SGR 1806-20
have been detected  (see \citet{Ersin2000ApJ,Prieskorn2012ApJ,bayrak2017ApJS} and references therein).
Detailed spectroscopic analyses were performed on some
bursts \citep{Atteia1987ApJ, Kouveliotou1987ApJ, Fenimore1994ApJ,bayrak2017ApJS}.
The observation data of the \textit{Rossi} X-ray Timing Explorer (RXTE)
show that the observed fluence ranges
between $\sim 10^{-10}$ and $ 10^{-7}$ erg cm$^{-2}$\citep{bayrak2017ApJS}.
Similarly, the fluence of the X-ray bursts detected by the International
Cometary Explorer (ICE) is in the range $\sim 10^{-8}$ --- $ 10^{-5}$ erg cm$^{-2}$
\citep{Atteia1987ApJ,Kouveliotou1987ApJ, Fenimore1994ApJ,Ersin2000ApJ}.
We notice that the fluence of the ICE bursts is higher than that of the
RXTE bursts by about two orders of magnitude. This may be due to the fact
that ICE has a much higher sensitivity, meaning that weaker bursts can
be recorded.

\begin{figure}
        \centering
        \includegraphics[width=0.95\columnwidth]{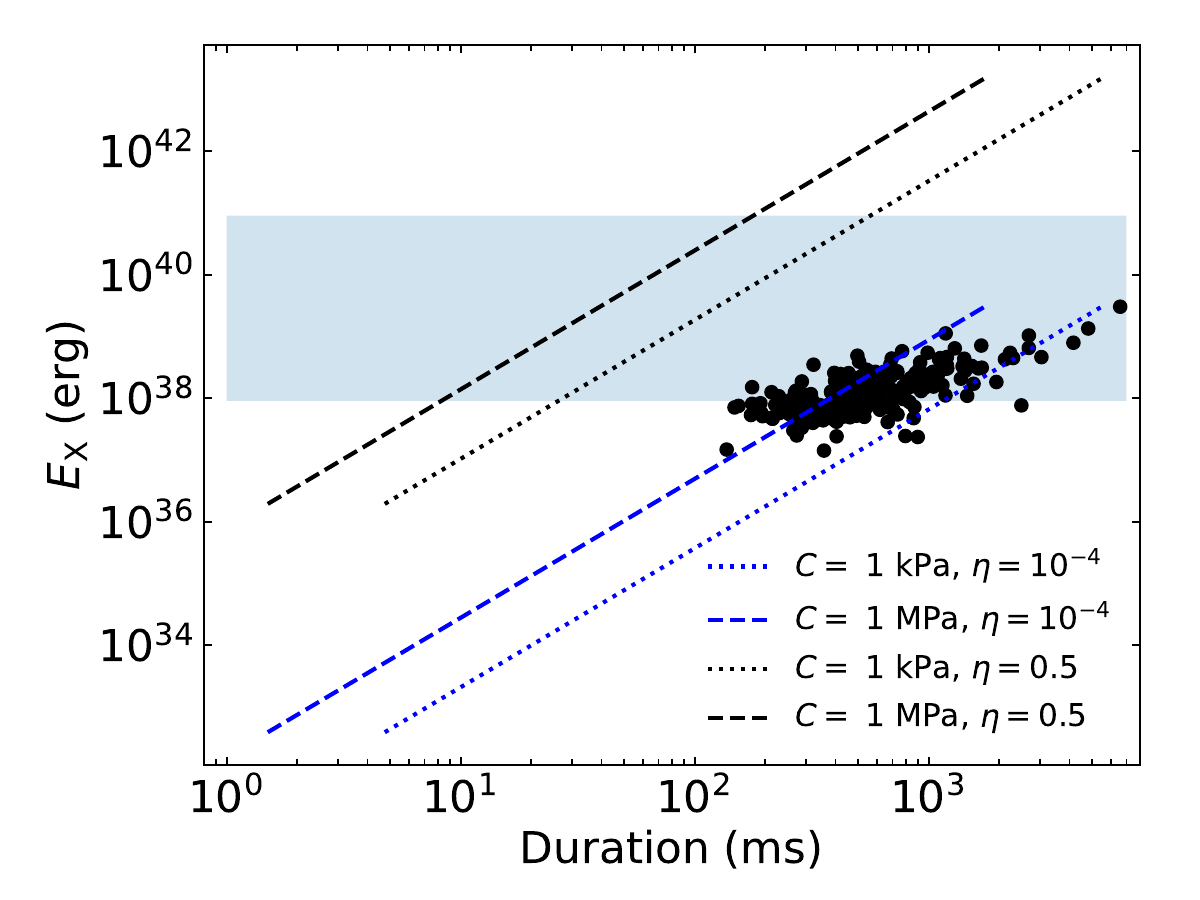}
        \caption{X-ray bursts from SGR 1806-20 plotted on the energy-duration
                plane. The black dots denote the observational data \citep{bayrak2017ApJS}.
                The lines are the
                theoretical results calculated by taking different values for the cohesive
                strength ($C$) and energy transforming efficiency ($\eta$).
                The shaded area represents the energy range of the bursts detected
                by the ICE detector \citep{Ersin2000ApJ}.
                \label{fig:1806}}
\end{figure}

Figure \ref{fig:1806} plots the burst energy as a function of the duration.
The isotropic energy of each burst is calculated by taking
the distance as 8.7 kpc \citep{Bibby2008MNRAS}. The data points correspond to the bursts
observed by RXTE, whose energy ranges between $\sim 10^{37}$ and $ 10^{39}$ erg.
The shaded area illustrates the energy range of the X-ray bursts detected by
ICE. We see that the energetics of the observed X-ray bursts can be explained well
 by our model.

Most of the X-ray bursts from SGR 1806-20 are concentrated near the
phase of $\sim$ 0.58 \citep{Zhang2021ApJ}. However,
the bursts could also appear at all the period phases.
In other words, the periodicity of SGR 1806-20 is not as strict as
that of SGR 1935+2154. The reason for such a relatively poor periodicity
may be that the system has a small $r_{\rm p}$, which can effectively
smear the periodicity to some extent (see Section \ref{sec:xrb}).

\section{Conclusions and discussion}
\label{sec:conclusion}

In this study, the periodicity and other properties of repeating X-ray bursts
from two magnetars are explained by the interaction between a
NS and its close-in planet. In our model, a rocky planet
moves around a NS in a highly eccentric orbit. The planet is partially disrupted
every time it passes through the periastron of the orbit due to the tidal force of
its compact host. During the process, clumps ranging from a few kilometers to $\sim$ 100
km are produced. The clumps can lose their angular momentum under the influence of
the remnant planet and fall toward the NS. They are further disrupted and captured
by the strong magnetic field when they enter the magnetosphere of the magnetar and/or NS.
The interaction between the NS and the clumps produces a series of X-ray bursts.
We show that the mass of the clumps that can enter the light cylinder radius is up
to $\sim10^{22}$ g, which is mainly determined by the clump's shearing strength. The
energy of the X-ray bursts produced in this way can be as high as $\sim10^{43}$ erg,
and their durations typically range from a few milliseconds to seconds. All these
features are consistent with observations. The bursts will show a clear
periodicity due to the orbital motion of the planet. The active window of the two
SGRs can also be explained \citep{Zhang2021ApJ, Zou2021ApJ, Xie2022MNRAS}.


It was reported that there were no bursts during some of the periods
\citep{Zhang2021ApJ, Zou2021ApJ}. In our framework, this can be caused by two factors. First, the asymmetry of mass loss can alter the
orbit. In the planet's subsequent passage through the periastron,
the mass loss would not occur if $r_{\rm p}$ were larger than the maximum
partial disruption distance ($r_{\rm td,max}$).
As a result, one would not be able to observe any bursts from the system.
Second, the Kozai-Lidov effect may be very complicated if the system
contains more than one planet.
Multi-planet systems have been identified in a few cases,
such as PSR 1257+12 \citep{Wolszczan1992Natur}, PSR B0943+10 \citep{Suleymanova2014ARep},
and PSR J1807-2459 \citep{Ransom2001ApJl, Ray2017ApJ}.
If the parameters of such systems satisfy some special conditions, rapid eccentricity
oscillation may occur due to the Kozai-Lidov effect, causing complicated
variations in the periastron distance ($r_{\rm p}$). There would be no mass
loss if $r_{\rm p} > r_{\rm td, max} $ during the oscillation process, and thus
no bursts would occur.

There is no clear correlation between the burst energy
and waiting time for the X-ray bursts from SGR 1806-20
\citep{Ersin2000ApJ}. This is consistent with the expectation
of self-organized criticality \citep{Katz1986JGR,Bak1987PhRvL}.
In our model, the waiting time is simply the interval time
between two successive collisions, which depends on the
spatial distribution of in-falling clumps. The breakup
distance of the clumps is generally determined by their
composition, but the breakup procedure itself is a random
process that leads to a stochastic distribution of the
size and in-fall time for the sub-clumps. As a result,
there is no correlation between the burst energy
and the waiting time in our framework, which agrees well with
the observations of SGR 1806-20.

For SGR 1806-20, a giant flare, whose
total energy release was $4.01\times10^{46}$ erg, was 
observed on 27 December 2004 \citep{Hurley2005Natur}.
Such an energetic flare is a rare event and one that had only been observed
from this source once before. If it were due to
a collision event, then the clump mass would be several
times $10^{26}$ g. In the partial tidal disruption process considered
here, the generation of such a massive clump
is possible \citep[e.g.,][]{Malamud2020a, Malamud2020b}.
Considering the collisions between clumps, the possibility of
some large objects being scattered toward the host star exists,
but the probability is low and it should be a rare
event \citep[e.g.,][]{Cordes2008ApJ}. If the clump is
composed of materials with a high cohesive strength, it can retain
a high mass before arriving at the light cylinder and will
eventually collide with the NS and/or SS to produce a giant flare
\citep[e.g.,][]{Zhang2000ApJ, Usov2001PhRvL}. Therefore, giant
flares are rare events in our framework, but they could reoccur in the future.

It is interesting to note that periodic X-ray flares are observed 
from Jupiter \citep{Yao2021sciadv}, which may be evidence 
that X-ray bursts can be generated via the interaction of external materials 
with a planet in the Solar System. It was argued that the charged 
ions that originate from the gas spewed into space due to giant volcano
events on Jupiter's moon (Io) could flow onto Jupiter along magnetic 
field lines, leading to an energy release in the form of X-rays\footnote{\url{https://phys.org/news/2021-07-scientists-year-mystery-jupiter-x-ray.html}}
\citep{Cowley2001PSS,Yao2021sciadv}.
Such phenomena could be universal and present across many
different environments in space \citep{Dunn2017NatAs, Yao2021sciadv, Mori2022NatAs}.
The X-ray bursts generated through the interaction of a NS with clumps disrupted
from a planet are to some extent similar phenomena.

Finally, we would like to mention that the dynamics of the clumps and 
the evolution of their angular momentum is highly complicated. 
Even in the asteroid-disruption explanation for the white dwarf pollution, 
it is still unclear how disrupted asteroids ultimately lose enough 
angular momentum to fall onto the white dwarf. There is a similar 
issue for the partial TDEs. Self-intersection
shocks arising from stream--stream collisions, which themselves
arise from relativistic apsidal precession, might play a role
in the dissipation. But this could be inefficient
in some circumstances. In our cases of partial disruption, the
remnant planet may act as a third object responsible 
for dissipation. Nevertheless, it is still possible that in some cases the
clumps may simply orbit around the NS many times, instead of
producing prompt accretion and X-ray bursts.

\begin{acknowledgements}
We would like to thank the anonymous referee for helpful suggestions 
that led to significant improvement of our study. 
This work was Sponsored by the Natural Science Foundation of Xinjiang Uygur Autonomous Region (No. 2022D01A363),
the National Natural Science Foundation of China (Grant Nos. 12033001, 12288102, 12273028 12041304, 12233002, 12041306),
the Natural Science Foundation of Xinjiang Uygur Autonomous Region (No. 2023D01E20),
the Major Science and Technology Program of Xinjiang Uygur Autonomous Region (Nos. 2022A03013-1, 2022A03013-3),
National Key R\&D Program of China (2021YFA0718500),
National SKA Program of China No. 2020SKA0120300,
the Tianshan Talents Training Program (Scientific and Technological Innovation Team).
YFH acknowledges the support from the Xinjiang Tianchi Program. AK acknowledges the support from the Tianchi Talents Project of Xinjiang Uygur Autonomous Region and the special research assistance project of the Chinese Academy of Sciences (CAS). This work was also partially supported by the Operation, Maintenance and Upgrading Fund for Astronomical Telescopes and Facility Instruments, budgeted from the Ministry of Finance of China (MOF) and administrated by the CAS, the Urumqi Nanshan Astronomy and Deep Space Exploration Observation and Research Station of Xinjiang (XJYWZ2303). 
\end{acknowledgements}

%
\bibliographystyle{aa} 
\bibliography{reference} 

\end{document}